\documentclass[final,5p,times,twocolumn]{elsarticle}

\usepackage{graphicx}
\usepackage{amsmath,amssymb}
\journal{Computer Physics Communications}

\newcommand{\pdagger}{{\phantom{\dagger}}}
\newcommand{\dt}{\Delta\tau}
\newcommand{\Dt}{\Delta\tau}
\newcommand{\reff}[1]{Fig.\ \ref{fig:#1}}

\newcommand{\refq}[1]{(\ref{eq:#1})}
\newcommand{\myparagraph}[1]{{\it #1} -- }

\begin{document}

\begin{frontmatter}

\title{Quantum Monte Carlo simulations of antiferromagnetism in ultracold fermions\\
on optical lattices within real-space dynamical mean-field theory}


%\author{}
\author{N.~Bl\"umer}
%\ead{Nils.Bluemer@uni-mainz.de}
\author{E.~V.~Gorelik}
%\affiliation{Institute of Physics, Johannes Gutenberg University, 55099 Mainz, Germany}
\address{Institute of Physics, Johannes Gutenberg University, 55099 Mainz, Germany}

\begin{abstract}
We present a massively parallel quantum Monte Carlo based implementation of 
real-space  dynamical mean-field theory for general inhomogeneous  
correlated fermionic lattice systems. 
As a first application, we study magnetic order in
a binary mixture of repulsively interacting fermionic atoms harmonically 
trapped in an optical lattice. We explore temperature effects
and establish signatures of the N\'{e}el transition in observables directly accessible in cold-atom experiments;
entropy estimates are also provided.
We demonstrate that the local density approximation (LDA) fails for ordered phases. 
In contrast, a ``slab'' approximation 
allows us to reach experimental system
sizes with ${\cal O}(10^5)$ atoms without significant loss of accuracy.
\end{abstract}


\end{frontmatter}

\label{Intro}
Starting with the achievement of Bose Einstein condensation, ultracold atomic gases have led to fascinating insights in quantum many-body phenomena \cite{Bloch08_RMP}. With the recent experimental successes in realizing quantum degenerate \cite{Koehl05} and strongly interacting Fermi gases \cite{Esslinger08_Nature,Bloch08_ferm} in optical lattices, such systems are considered as highly tunable {\it quantum simulators} of condensed matter \cite{Bloch08_Science}.
The recent observation of the fermionic Mott transition in binary mixtures of $^{40}K$ on %three-dimensional 
cubic lattices \cite{Esslinger08_Nature,Bloch08_ferm} marks important progress in this respect.
The next experimental challenge, taken up by many of the leading groups, is to reach and identify an ordered antiferromagnetic (AF) N\'{e}el phase.

Ultracold quantum gases on optical lattices have several advantages 
in comparison to solid state systems; however, the detection methods established
for solids do not always find correspondence in cold-atom experiments. In particular, the detection of
the AF order parameter
is not straightforward. Therefore, it is important to identify fingerprints of AF phases that are directly accessible 
experimentally.

Due to the intrinsic inhomogeneity of trapped atomic clouds, even the qualitative interpretation of experimental data 
may depend on corresponding quantitative simulations. The dynamical mean-field theory (DMFT) is well established 
as a powerful, nonperturbative approach to interacting Fermi systems \cite{Georges96,Kotliar_Vollhardt}. 
 Spatial inhomogeneities of the optical lattice can be captured within a real-space 
extension of the method (RDMFT) \cite{Snoek_NJP08,Helmes_PRL08} 
or, approximately, by applying DMFT within a local density approximation (LDA).
In either case, the numerical accuracy depends on the method 
chosen as DMFT impurity solver. Quantum Monte Carlo (QMC) based solvers are precise
and even numerically cheap at or above the temperature ($T$) ranges relevant for AF ordering and accessible
in cold-atom experiments.

In this paper, we introduce our Hirsch-Fye QMC \cite{Hirsch86} based RDMFT implementation, with
focus on computational aspects.
Then, we turn
to a first application, establishing the essential physics related to AF order in a relatively small system involving
${\cal O}(100)$ atoms. In order to minimize finite-size effects, we choose a square lattice geometry
for which finite-temperature antiferromagnetism is not really physical; however, due to the mean-field
character of the DMFT, the results are representative of three-dimensional cubic systems, up to 
a change in energy scales. Using selective simulations with ${\cal O}(1000)$ atoms, we establish
that the properties of large three-dimensional clouds can be accurately extracted from simulations for
central two-dimensional slabs and that simulation boxes can be chosen smaller than one would naively
expect.

%%%%%%%%%%%%%%%%%%%%%%%%%%%%%%%%%%%%%%%%%%%%%%%%%%%%%%%%%%%%%%%%%%%%%%%%%%%%%%%%%
%%%%%%%%%%%%%%%%%%%%%%%%%%%%%%%%%%%%%%%%%%%%%%%%%%%%%%%%%%%%%%%%%%%%%%%%%%%%%%%%%

%\section{Model and Methods}
\myparagraph{Model and Methods}
Binary mixtures of equivalent fermions in an optical lattice are well described by the Hubbard
model [with (isotropic) trapping potential $V_i = V_0 r_i^2/a^2$],
  \begin{equation}\label{Hubb_mod}
    \hat{H} =\! -t \sum_{\langle ij\rangle ,\sigma}  \hat{c}^{\dag}_{i\sigma}
  \hat{c}^\pdagger_{j\sigma} 
  + \! U \sum_{i} \hat{n}_{i\uparrow} \hat{n}_{i\downarrow}
  \,+ \sum_{i, \sigma}(V_i - \mu)\, \hat{n}_{i\sigma}.
  \end{equation}
  Here, $\hat{n}_{i\sigma} = \hat{c}^{\dag}_{i\sigma} \hat{c}^\pdagger_{i\sigma}$, 
$\hat{c}^\pdagger_{i\sigma}$ ($\hat{c}^{\dag}_{i\sigma}$) are annihilation (creation)
operators for a fermion with (pseudo) spin $\sigma\in \{\uparrow,\downarrow\}$ at site $i$ (with coordinates ${\bf r}_i$), 
$t$ is the hopping amplitude between nearest-neighbor sites $\langle ij\rangle$, $U > 0$
is the on-site interaction, and $\mu$ is the chemical potential.
We choose $V_0=0.25t$ and use $t$ as the energy unit.

%%%%%%%%%%%%%%%%%%%%%%%%%%%%%%%%%%%%%%%%%%%%%%%%%%%%%%%%%%%%%%%%%%%%%%%%%%%%%%%%%
\begin{figure} 
	\includegraphics[width=\columnwidth]{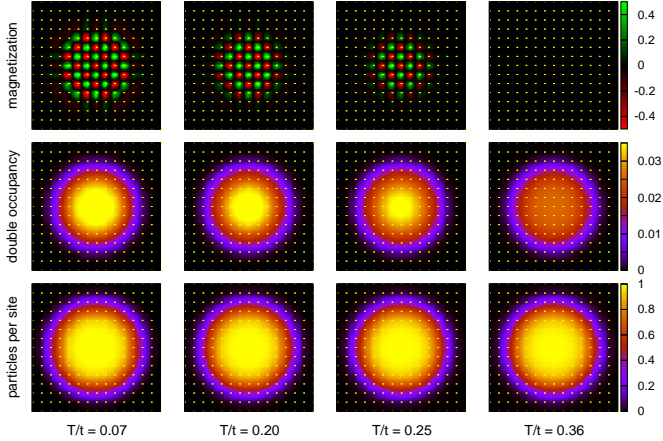}
\caption{RDMFT-QMC results for square lattice ($V_0/t=0.25$, $U/t=10$). 
First row: as evidenced by the local magnetization, 
a large AF core is strongly polarized (up to 90\%) for $T\lesssim 0.14t$ (left column);
with increasing $T$, both extent and magnitude of the AF order decay until a paramagnetic
phase is reached for $T>T_{\text{N}} \approx 0.32t$.
Second row: the double occupancy is strongly enhanced in AF regions; in contrast, the density (third row) 
shows little $T$ dependence. }
\label{fig:melting} 
\end{figure}
%%%%%%%%%%%%%%%%%%%%%%%%%%%%%%%%%%%%%%%%%%%%%%%%%%%%%%%%%%%%%%%%%%%%%%%%%%%%%%%%%

The RDMFT approach \cite{Snoek_NJP08,Helmes_PRL08} is based on the following
expression for the Green function matrix $G$ for Matsubara frequency
$i\omega_n=(2n+1)i\pi T$ (with spin indices suppressed):
\begin{equation}\label{eq:RDMFT}
\big[{G}({i\omega_n})\big]^{-1}_{ij} \! = t_{ij} + \big[{i\omega_n} + {\mu} - {V_i} - {\Sigma_{i}}({i\omega_n})\big]\, \delta_{ij}\,; 
\end{equation}
here the only approximation is the DMFT assumption of a local (i.e. site-diagonal) self-energy $\Sigma$,
which corresponds to a momentum independence of $\Sigma$ in translation-invariant systems \cite{Kotliar_Vollhardt}.
The standard DMFT impurity problem, one for each inequivalent lattice site, is solved in this work using the Hirsch-Fye QMC algorithm \cite{Hirsch86}. The discretization of the imaginary-time interval $0\le \tau \le 1/(k_B T)$ introduces a systematic error which can be eliminated by extrapolations $\dt\to 0$ \cite{Bluemer05,Bluemer07}. In this work, we choose $\Dt \,t=0.1$ unless indicated otherwise.

Note that \refq{RDMFT} only couples $G$ and $\Sigma$ at the same Matsubara frequency; thus, the matrix inversions can be performed in parallel for the 500 Matsubara frequencies explicitly taken into account. In contrast, the impurity equations couple different frequencies (and imaginary times), but are local in the site indices. Therefore, this part
of the self-consistency problem can be performed in parallel for all inequivalent impurities. In addition, the importance sampling for each impurity may be distributed (via MPI) over some $10-100$ processes.
In total, the program can saturate 500 CPU cores for large enough problems.

%%%%%%%%%%%%%%%%%%%%%%%%%%%%%%%%%%%%%%%%%%%%%%%%%%%%%%%%%%%%%%%%%%%%%%%%%%%%%%%%%
%%%%%%%%%%%%%%%%%%%%%%%%%%%%%%%%%%%%%%%%%%%%%%%%%%%%%%%%%%%%%%%%%%%%%%%%%%%%%%%%%

%%%%%%%%%%%%%%%%%%%%%%%%%%%%%%%%%%%%%%%%%%%%%%%%%%%%%%%%%%%%%%%%%%%%%%%%%%%%%%%%%
\begin{figure} 
	\includegraphics[width=\columnwidth]{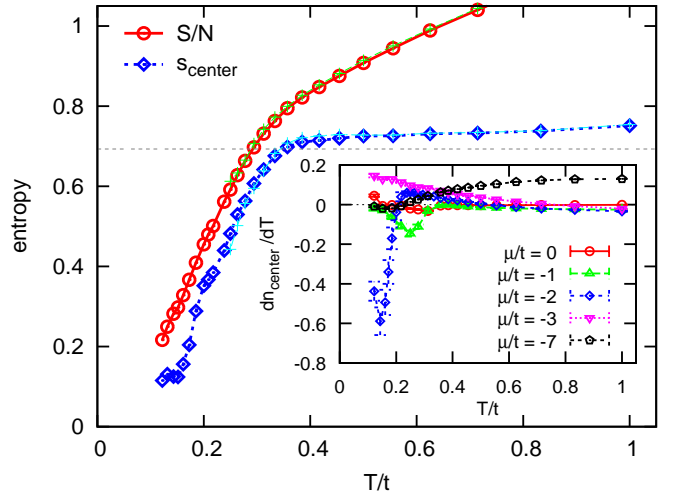}
\caption{Main panel: RDMFT-QMC estimates of entropy for square lattice ($V_0/t=0.25$, $U/t=10$);
QMC data for the average entropy per particle (circles, for discretization $\Dt=0.1$) and
for the entropy of the central site (squares, also for $\Dt=0.1$) are hardly distinguishable from
corresponding estimates for $\Dt=0.05$ (crosses).
Inset: temperature derivative $dn_{\text{central}}/dT$ of the central density
for representative values of the chemical potential $\mu$.} 
\label{fig:entropy} 
\end{figure}
%%%%%%%%%%%%%%%%%%%%%%%%%%%%%%%%%%%%%%%%%%%%%%%%%%%%%%%%%%%%%%%%%%%%%%%%%%%%%%%%%

\myparagraph{Results for square lattice}
For the square lattice, the noninteracting dispersion extends from $-4t$ to $4t$, leading to a bandwidth of $W=8t$. The chosen interaction $U=10t$ is already in the strongly correlated regime: at half filling, the fermions would always be localized. However, in the inhomogeneous trapped system, an outer shell has to remain delocalized due to the outward decay of the filling to zero. Consequently, staggered magnetic order can only be expected in the core region.

This is exactly what is seen in \reff{melting} (first row): at low $T$ (left column), the core shows a nearly perfect staggered magnetization. With increasing $T$ (from left to right), both the extent of the ordered region and its polarization decrease until the order is lost at the N\'{e}el temperature $T_{\text{N}} \approx 0.32t$. Unfortunately, this most obvious signature of AF order is not directly accessible experimentally, primarily due to lack of single-site resolution. 
A measurement of particle density profiles (third row) also wouldn't help in this case as they are practically unchanged (at this scale) through the transition. In contrast, the double occupancy, the probability of two particles occupying the same lattice site, provides a distinct signal: at high $T$, it is featureless in the center, with a maximum value of about $0.02$. Only at low temperatures, it is enhanced, by up to $50\%$, in the developing central antiferromagnetic core. For measurements of the double occupancy, accuracies of about $0.01$ have already been established \cite{Esslinger08_Nature}; thus, only a minor refinement is needed to resolve this AF signal (when low enough $T$ can be reached).

Experimentally, the fermionic systems are prepared without an optical lattice and with negligible interactions; 
the initial temperature can, therefore, be determined by fitting the density profile with a Fermi distribution function. Then, the lattice is switched on; due to the localization, this also induces a contact interaction.
Clearly, the temperature of a final equilibrium state will differ greatly from the initial value (in an unknown way); however, the process is essentially adiabatic. Therefore, only the entropy density is a reliable temperature scale.

%%%%%%%%%%%%%%%%%%%%%%%%%%%%%%%%%%%%%%%%%%%%%%%%%%%%%%%%%%%%%%%%%%%%%%%%%%%%%%%%%
\begin{figure} 
	\includegraphics[width=\columnwidth]{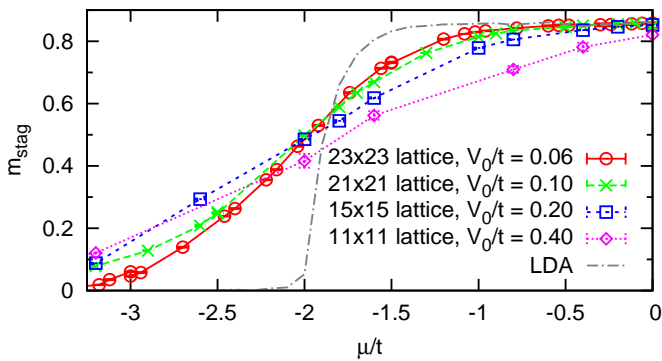}
\caption{RDMFT-QMC estimates of the AF order parameter, i.e., the staggered magnetization, for square lattice ($U/t=8$, $T/t=0.14$, half filling at central site) for various trap stengths $V$ (symbols) as a function of the effective chemical potential $\mu_i=\mu_0-V_0 r_i^2/a^2$. Even for nearly realistically weak trap strength ($V_0=0.06$), the transition is much smoother, by proximity effects, than estimated from LDA (using single-site DMFT and QMC; dash-dotted line).} 
\label{fig:proximity} 
\end{figure}
%%%%%%%%%%%%%%%%%%%%%%%%%%%%%%%%%%%%%%%%%%%%%%%%%%%%%%%%%%%%%%%%%%%%%%%%%%%%%%%%%

Monte Carlo importance sampling cannot directly access entropy (or free energy) information. Thus, we revert to the thermodynamic relation $dS/d\mu=dN/dT$ and obtain the entropy density from integrating over temperature derivatives of particle densities (using the fact that $\mu\to-\infty$ corresponds to an empty system with zero entropy). The result is shown in \reff{entropy}: after a steep initial increase, the {\em central} entropy density (diamonds) develops a kink around $T_{\text{N}}\approx 0.32t$ (arrow) and then reaches a plateau with a value slightly above the mean field value of $\log(2)$ for a spin system. The {\em average} entropy density (circles) is always larger due to the low-density shell; in particular, the kink is much weaker and a significant slope remains above $T_N$. Corresponding entropy data for smaller discretization (crosses) show excellent agreement, which establishes that systematic errors are insignificant. We should point out that fine grids in $T$ and $\mu$ are essential for this study since strong peaks in the integrand $dn/dT$ (see inset of \reff{entropy}) 
%close to phase boundaries 
have to be resolved.

One might ask whether expensive RDMFT calculations as described above are really necessary for properly resolving the ordering phenomena. After all, in the case of the paramagnetic Mott metal-insulator transition, the LDA, which approximates the properties of each site by those of a homogeneous system with the same local potential, yields static observables in nearly perfect agreement with RDMFT \cite{Helmes_PRL08}. However, as seen in \reff{proximity}, the order parameter profiles show enormous deviations from LDA even for the largest systems, which fully justifies the larger numerical effort of the RDMFT approach.

\myparagraph{Efficient simulations for three dimensions} Let us, finally, leave the ``toy'' case of small two-dimensional systems and discuss strategies for realistic simulations of current experimental setups involving cubic lattices and ${\cal O}(10^5)$ particles. While such system sizes do not pose fundamental difficulties on the impurity part of the self-consistency problem, due to linear scaling and perfect parallelism, a full matrix inversion (with cubic scaling in the number of sites) as required in \refq{RDMFT} is out of question, at least for dense-matrix algorithms. In our current implementation and for typical numerical accuracies, the cost of the matrix inversion dominates beyond some 1000 lattice sites.

Fortunately, the system sizes that have to be simulated explicitly can be reduced quite drastically.
First of all, periodic boundary conditions should be used even for the inhomogeneous trapped systems. Then, 
the simulation box can be chosen much smaller than the atomic cloud provided that the AF core \cite{fn:isotropy} does not touch the boundaries, as shown in \reff{FS_slab}: while spikes are visible (at radial distances of about $7a$) in $m_{\text{stag}}$ for a $14\times 14\times 14$ system (lower panel, short-dashed lines) at $T/t=0.33$, the $16\times 16\times 16$ data (long-dashed lines) are already smooth at this scale, indicating that finite-size effects become negligible.
At $T/t=0.41$, closer to the N\'eel temperature $T_{\text{N}}\approx 0.45t$, the accuracy in $m_{\text{stag}}$ improves further due to the smaller core. 

Even larger savings are possible by restricting the simulation to a central slab, with periodic boundary conditions in the perpendicular direction. In practice, we neglect the (small) perpendicular component of the trapping potential in the slab, i.e., replace the spherical potential by a cylindrical one, which is an excellent approximation for large enough systems. Then, a thermodynamic limit is reached, within numerical accuracy, already at layer thickness of 4. Corresponding results, indicated by solid and dotted lines in \reff{FS_slab}, show negligible size dependence and agree well with the direct 3D results, independent of the temperature (density profiles for $T/t=0.41$, nearly indistinguishable from  those for $T/t=0.33$, are omitted). In contrast, the LDA (dash-dotted lines) is completely off for $m_{\text{stag}}$.

%%%%%%%%%%%%%%%%%%%%%%%%%%%%%%%%%%%%%%%%%%%%%%%%%%%%%%%%%%%%%%%%%%%%%%%%%%%%%%%%%
\begin{figure} 
	\unitlength0.1\columnwidth
	\begin{picture}(10,8)
    \put(0,0){\includegraphics[width=\columnwidth]{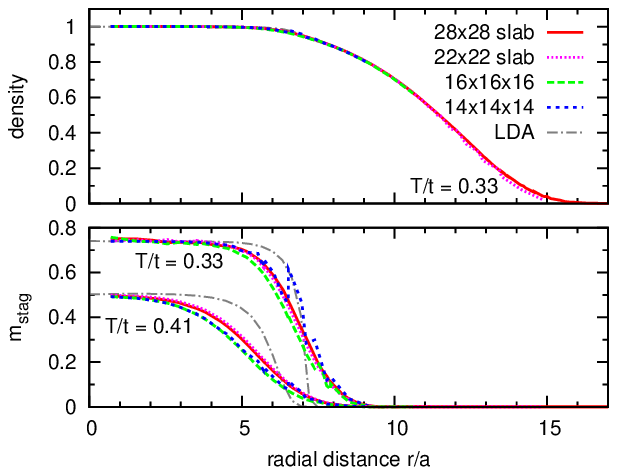}}
    \put(6.0,1.1){\includegraphics[width=0.36\columnwidth]{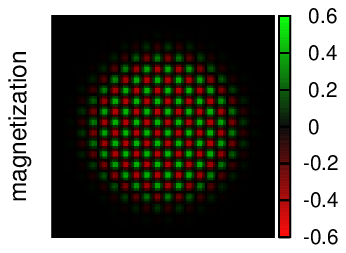}}
    \put(1.7,4.3){\includegraphics[width=0.36\columnwidth]{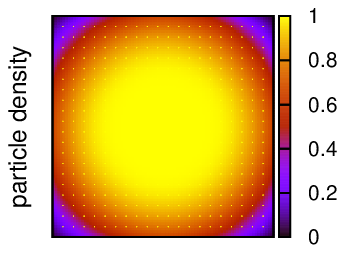}}
  \end{picture}
  \caption{RDMFT-QMC estimates for cubic lattice ($V_0/t=0.05$, $U/t=12$) using periodic boundary conditions, either for a full 3D lattice or using the slab approximation (see text). For the density (top panel), the slab results (dashed lines) are hardly distinguishable from the full calculations (solid and dotted lines); finite-size effects are very small. The agreement is also very good for the order parameter (bottom panel), even close to the N\'eel temperature $T_{\text{N}}\approx 0.45t$, with the exception of spikes corresponding to edges of the 3D systems. Insets: central density and AF ordering pattern from $22\times 22$ slab calculation.} 
\label{fig:FS_slab} 
\end{figure}

In conclusion, we have implemented the RDMFT approach for inhomogeneous correlated Fermi systems
on a QMC basis. 
In this work, we have shown, for a relatively small system, that the onset of antiferromagnetic 
order at low $T$ is signaled by an enhanced double occupancy. This effect and the impact of
the interaction strenght $U$ are discussed on a more quantitative level for large cubic systems
elsewhere \cite{Gorelik10}, using the slab approximation established in this paper.

We thank W.\ Hofstetter, M.\ Snoek, and I.~Titvinidze for valuable discussions and help in implementing RDMFT.
Support by the DFG within the TR 49 and by the John von Neumann Institute for Computing is
gratefully acknowledged.

%\vspace{-3ex}

\end{document}